\documentstyle[12pt]{article}
\begin{document}

\def\p{\phi}
\def\P{\Phi}
\def\a{\alpha}
\def\e{\varepsilon}
\def\be{\begin{equation}}
\def\ee{\end{equation}}
\def\l{\label}
\def\0{\setcounter{equation}{0}}
\def\T{\hat{T}_}
\def\b{\beta}
\def\S{\Sigma}
\def\3{d^3{\rm \bf x}}
\def\4{d^4}
\def\C{\cite}
\def\r{\ref}
\def\ba{\begin{eqnarray}}
\def\ea{\end{eqnarray}}
\def\nn{\nonumber}
\def\R{\right}
\def\L{\left}
\def\q{\hat{Q}_0}
\def\X{\Xi}
\def\x{\xi}
\def\la{\lambda}
\def\d{\delta}
\def\s{\sigma}
\def\f{\frac}
\def\vx{{\rm \bf x}}
\def\j{\frac{\delta}{i \delta j_a ({\rm \bf x},x_0+t+t_1)}}

\begin{titlepage}

\begin{flushright}
{\normalsize IP GAS-HE-21/95}
\end{flushright}
\vskip 3cm

\begin{center}
{\Large \bf Comments to the S-matrix interpretation of finite-temperature
field theories}
\vskip 1cm

\mbox{J.Manjavidze}
\footnote {Institute of Physics, Georgian Academy of Science,
Tamarashvili st.6, Tbilisi, 380077, Republic of Georgia,
e-mail: jm@physics.iberiapac.ge}\\
\end{center}
\vskip 1.5cm

\begin{abstract}
\footnotesize
It is shown that there is the possibility to find at least in the
perturbation framework the Matsubara theory from the $S$-matrix
interpretation of the real-time finite-temperature field theory
if the system under consideration is in an equilibrium state.
\end{abstract}
\end{titlepage}

\section{Introduction}
\setcounter{equation}{0}

In this paper we will discuss the $S$-matrix interpretation of the
real-time finite temperature field theory \C{1} in connection with
the Matsubara imaginary time theory \C{mats}. We shell show the
qualitative conditions in frame of which both theories are
coinside.

It will be seen that the main condition is the state of equilibrium.
We will define the equilibrium state as a state with Gaussian fluctuations
of the thermodynamical parameters. This is evident for Matsubara
theory. But in the $S$-matrix formalism it gives the nontrivial
constraints since such quantity as, for instnce, the temperature
introduced in it as the Lagrange multiplier, so as in the
microcanonical formalism.

We will assume also that the perturbation theory is applicable. The
importance of this condition will be discussed in Sec.3.

It will be shown in result that under this conditions the $S$-matrix
theory can be analyticaly continued to the Matsubara imaginary time theory
(Sec.2). This result seems important since it shows up the ability of the
$S$-matrix approach to describe the statistical systems.

The introduction of the temperature $T$ as the Lagrange multiplier
means that the parameter $T$ introduced in a theory for sake of
``economy" description of the many-particle system. In Sec.3
we find the quantitative conditions when such symplification
is valid. We will show that $T$ is a ``good" parameter if
there is asymptotic connections between veriouse energy
correlation functions.

\section{The S-matrix finite-temperature field theory}
\setcounter{equation}{0}

The experience of papers \C{1,2} shows that the Wigner functions
\C{wig} generating functional \C{car} $R_S$ can be written in
the $S$-matrix framework in a factorised form independently from
boundary conditions:
\be
R_S(\b_+ ,\b_- )=
e^{N(\hat{\phi}^*_i\hat{\phi}_j; \b_+ ,\b_- )}
R_0 (\phi_{\pm}),
\l{1}
\ee
where $N$ is some nonlinear operator of
\be
\hat{\phi}_i (q)=
\int dx e^{-iqx} \delta /\delta \phi_i (x), i=+,-.
\l{}
\ee
The concrete form of this operator depends from boundary conditions,
i.e. from the environment of the system. For the case of uniform
temperature distributuin,
\ba
N(\hat{\phi}^*_i \hat{\phi}_j ;\b_+ ,\b_-)=
\int d\omega (q) \hat{\phi}^*_i (q) n_{ij}(q,(\b))\hat{\phi}_j (q)+...,
\nn\\
d\omega (q)=\frac{d^3 q}{(2\pi )^3 2\epsilon (q)},
\epsilon (q)=(q^2 +m^2 )^{1/2}.
\l{2}
\ea
(The summation over all configurations of $(ij)$ is assumed.) Here
$n_{ij} (q)$ is the occupation number. Assuming that the environment is
composed from noncorrelated particles one can calculate \C{1} that
\ba
n_{++}=n_{--}=\tilde{n}(|q_0|(\b_+ +\b_- )/2),
\nn\\
n_{+-}=\Theta (q_0)(1+\tilde{n}(q_0 \b_+))+
\Theta(-q_0)\tilde{n}(-q_0\b_-),
\nn\\
n_{-+}=\Theta(q_0)\tilde{n}(q_0 \b_+)+\Theta (-q_0)
(1+\tilde{n} (-q_0 \b_-)),
\l{}
\ea
where
\be
\tilde{n}(q_0\b)=\frac{1}{e^{\b q_0}-1}, q_0 =(q^2 +m^2)^{1/2}
\l{}
\ee
is the mean multiplicity of environment purticles with energy
$q_0$ at a given temperature $1/\b$. The subsecuent terms in
(\ref{2}) describes the correlations among particles of the
environment. In this formulation of the theory they are the free
parameters. We can consider only the first term assuming absence of
correlations.

We shell interpret $1/\b_-$ as the temperature of initial state and
$1/\b_+$ as the final state temperature \C{1}. Therefore, by definition,
the $S$-matrix theory is the two-temperature.

The generating functional
\be
R_0 (\phi_{\pm})=Z(\phi_+)Z(\phi_-),
\l{}
\ee
where $Z(\phi_{\pm})$ are the vacuum into vacuum transition amplitudes:
\be
Z(\phi_{\pm})=\int D\Phi_{\pm}e^{\pm iS_0 (\Phi_{\pm})\mp
iV(\Phi_{\pm}+\phi_{\pm})}.
\l{3}
\ee
They were defined in \C{1} on the Mills' time contour
\C{mil}:
\be
\Phi_{\pm}=\Phi_{\pm} (t \in C_{\pm}),
C_{\pm}:t \rightarrow t \pm i\e , \e \rightarrow +0,
-\infty \leq t \leq +\infty.
\l{}
\ee
In eq.(\ref{3}) $S_0$ is the free part of the action and $V$ describes the
interactions. Calculating integral (\ref{3}) the usual field-theoretical
boundary conditions:
\be
\int_{\sigma_{\infty}} d\sigma_{\mu}
\Phi_{\pm}\partial^{\mu}\Phi_{\pm}=0
\l{}
\ee
must be applied since the environment is fixed by the operator $N$.
It means that ($\sigma_{\infty}$ is the infinitly far hypersurface)
\be
\Phi_{\pm} (t =-\infty)=0, \Phi_{\pm} (t =+\infty)=0
\l{12}
\ee
(the space boundary conditions are trivial). In the perturbation theory
framework
\be
Z(\phi_+)=
e^{-i\int dx \hat{j}_+ (x)\hat{\Phi}_+ (x)}
e^{-iV(\Phi_+ +\phi_+}
e^{-\frac{i}{2}\int dx dx' j_+ (x)D_{++}(x-x')j_+ (x')},
\l{4}
\ee
where $D_{++}(x-x')$ is the causal (Feynman) Green function and
$\hat{j}(x) \equiv \delta /\delta j(x)$,
$\hat{\Phi}(x) \equiv \delta /\delta \Phi(x)$.

So, we see that $R_0 (\phi_{\pm})$ describes pure fields dynamics and
$N(\hat{\phi}^*_i\hat{\phi}_j ;\b_+ ,\b_-)$ containes the
thermodynamics. One can say that the action of operator $\exp\{N\}$
maps the interacting fields asystem on the thermodynamical state.

Using (\ref{1}), (\ref{2}) and (\ref{4}) we can find that
\be
R_S (\b_+ ,\b_- )=
e^{-iV(-i\hat{j}_+)+iV(-i\hat{j}_-)}
e^{\frac{i}{2}\int dx dx' j_i (x)G_{ij}(x-x',(\b))j_j (x')},
\l{5}
\ee
where the $n_{ij}$ dependence was absorbed in the Green functions
$G_{ij}$. In the momentum represantation the Green functions
looks as follows:
\ba
i\tilde{G}(q,(\b ))=
\left (
\matrix{
\frac{i}{q^2 -m^2 -i\e}& 0 \cr
0 & -\frac{i}{q^2 -m^2 -i\e} \cr
} \right )
+\nn \\ \nn \\ +
2\pi \d (q^2 -m^2 )
\left (
\matrix{
\tilde{n}(\frac{\b_+ +\b_-}{2} |q_0|)&
\tilde{n}(\b_+ |q_0|)a_+ (\b_+) \cr
\tilde{n}(\b_- |q_0|)a_- (\b_-) &
\tilde{n}(\frac{\b_+ +\b_-}{2}|q_0| \cr
} \right ),
\l{6}
\ea
and
\be
a_{\pm}(\b )=-e^{\b (|q_0| \pm q_0 )/2}
\l{}
\ee
The matrix Green functions was introduced firstly in \C{moh}.

Let us assume now that the energy $E$ of the system is fixed and,
therefore, $\b_{\pm}$ are the flactuating quantities. The corresponding
generating functional $r(E)$ can be deduced calculating the integrals
\C{1}:
\be
r(E)=\int \frac{d\b_+}{2\pi i}\frac{d\b_-}{2\pi i}
e^{E(\b_+ +\b_-)-F(\b_+ ,\b_- )},
\l{7}
\ee
where
\be
F(\b_+ ,\b_-) \equiv -\ln R_S(\b_+ ,\b_-)
\l{15}
\ee
and the integrations are performed along the imaginary axis. This integrals
we shell compute by the stationary phase method. In our case there are
two equations of state (we destinguish the initial and final states
temperatures):
\be
E=\frac{\partial}{\partial \b_i}F(\b_+ ,\b_-), i=+,-.
\l{}
\ee
In absence of the energy dissipation this two equations have the
same solution:
\be
\b_{\pm}=\b (E)>0, Im \b =0.
\l{}
\ee
Expanding integrals in (\ref{7}) over $(\b_+ -\b)$, $(\b_- -\b)$ we
shell leave onle first term. This imply that the fluctuations of
$\b_+$ and $\b_-$ near $\b (E)$ are Gaussian. This question will be
discussed also in Sec.3.

In result, the generating functional in the energy represantation is
\be
r(E) \sim R_S(\b_+ =\b,\b_- =\b)| det
\left ( \frac{\partial^2}{\partial \b_i\partial \b_j }
F(\b_+ ,\b_-)\right) \|_{\b_{\pm}=\b}|^{-1/2}
e^{2\b (E)E}.
\l{}
\ee
It was assumed that
\be
det \left( \frac{\partial^2}{\partial \b_i\partial \b_j }
F(\b_+ ,\b_-) \right) \neq 0.
\l{}
\ee
So, the generating functional in the temperature represantation
for equilibrium case is
\be
R_S(\b)=R_S(\b_+ =\b,\b_- =\b).
\l{}
\ee
We can compare now $R_S (\b)$, defined in (\ref{5}), with the
Niemi-Semenoff's generating functional \C{sem} $R_{NS}$.

The path integral represantation for generating functional $R_{NS}$
has the form \C{sem}:
\be
R_{NS}(\b)=\int D\Phi e^{iS_{C_{\b}(T_-,T_+)}(\Phi ,j)}
\l{8}
\ee
where the total action includes an external source $j$:
\be
S_{C_{\b}(T_-,T_+)}(\Phi ,j)=S_{C_{\b}(T_-,T_+)}(\Phi )+
\int_{C_{\b}(T_-,T_+)} dx j(x) \Phi (x).
\l{9}
\ee
It is remarkable that the integration performed in (\ref{8}) over
fields defined on the Mills' closed-time contour. So,
\ba
C_{\b}(T_-,T_+)=C_{1}(T_-,T_+)+C_{2}(T_+,T_+ -i(\b +\a)/2)+
\nn \\
+C_{3}(T_+ -i(\b +\a )/2,T_- -i(\b +\a )/2)
+C_{4}(T_- -i(\b +\a )/2,T_- -i\b ).
\l{10}
\ea
This meance that the time integral in (\ref{9}) start from the
point $T_- +i\e$, $\e \rightarrow +0$, and goes to $T_+ +i\e$,
$T_+ > T_-$. It is the $C_1$ part of $C_{\b}$. Contour $C_2$
start from $T_+ +i\e$ and goes to $T_+ -i(\b +\a)/2$, etc. The
times $T_-$, $T_+$ can be choosen arbitrary. So, if $T_+ =T_-$
the contour $C_{\b}(T_-,T_-)$ will coinside with Matsubara
time contour \C{mats}.

Note also that the theory has the remarkable degree of freedom \C{*}
which follows from the translational invariance of the system. One
can choose arbitrary $\a$ in the interval:
\be
-\b  \leq \a \leq \b.
\l{}
\ee
This gives the possibility to assume that there is not singularities
in the strip $(0,-i\b )$ of the complex time plane. Last one formally
allows to introduce the Kubo-Martin-Schwinger (KMS) \C{kms} boundary
condition:
\be
\Phi (T_- )=\Phi (T_- -i\b ).
\l{11}
\ee
This is the last ingredient of the Schwinger's real-time formalis
\C{sch}. Note that the temperature introduced through the boundary
condition (\ref{11}).

We can consider the limit $T_{\pm} \rightarrow \pm \infty$. In the
perturbation framework the contributions from contours $C_2$ and
$C_4$ are disappear in this limit and choosing $\a =-(\e +\b )$
we will find the represantation (\ref{5}) from (\ref{8}).

This demonstrates the equivalence of the $S$-matrix theory and of
the traditional approach to the thermodynamics if (i) the
perturbation theory is used and if (ii) the infinite time interval
($T_+ =+\infty$) is considered. Last one is important since only in
this limit we are able to see that addition of $C_2$ contour do not
give a new contribution.

In the $S$-matrix formalism we had $two$ boundary conditions (\ref{12}).
The second boundary condition of (\ref{12}) should be changed on the contour
$C_2$ in the closed-time path formalism. In result fields are defined
on the contour $C_{\b}(T_-,T_+)$.

Therefore, under above mentioned constraints we find the equality:
\be
R_S (\b )=R_{NS} (\b )|_{T_{\pm}=\pm \infty},
\l{13}
\ee
So, $R_S (\b )$ can be written in the ``closed-time path" form (\ref{8}),
i.e. can be defined on the Matsubara imaginary-time contour.

\section{Concluding remarks}
\setcounter{equation}{0}

Now it is useful to chek up the constraints under which the equality
(\ref{13}) is valid.

A. {\it Factorzation}.

The eq.(\ref{1}) fixes the statement that one can find the operator
which maps the ``mechanical" system of interacting fields on the
thermodynamical state. The eq.(\ref{1}) was derived in \C{1} adopting the
reduction formulae to the path-integral formalism. Noting presence of
the reduction formulae criticism it must be mentioned that this
formulae works well at least in the perturbation theory.

It must be mentioned also that we can introduce the $S$-matrix as the
assumption and then to check this assumption at the very end of
calculations. One can find, for instance, that the (hidden)
conservation laws lead to the trivial generating functional
$R_S=R_0 (0)$. This meance that the auxiliary (external) field $\phi$
can not interact with field $\Phi$ through the potential
$V(\Phi +\phi )$, see (\ref{3}) (the field $\Phi$ is ``confined"
in this case).

The $S$-matrix approach gives a possibility to investigate this
solution. It is interesting to find this possibility in the
``closed-time path" formalism also (the problem is connected
with boundary condition (\ref{11}) since it hiddely assumes
that the field $\Phi$ is not confined: $1/\b$ is the
temperature of the interacting fields $\Phi$ system).

B. {\it Absence of correlations in the environment}.

We had leave only first term in the expansion (\ref{2}) of operator
$N$. If there is the correlations among particles of the environment
the higher powers of $(\hat{\phi}^*_i\hat{\phi}_j)$ must be taken into
account. In this case the theory will contain arbitrary number of
phenomenological parameters. The analogiouse realisation of Schwinger's
formalism was offered in \C{hu}. But there is the difference between
$S$-matrix and this generalized approache.

In the $S$-matrix formalism the operator $N$ containes the only product
$(\hat{\phi}^*_i\hat{\phi}_j)$. This allows to interpret
$(\hat{\phi}^*_i\hat{\phi}_j)$ as the operator of particles number.
Indeed, $\hat{\phi}^*_i$ is the creation operator and $\hat{\phi}_i$
is the absorbtion one. Therefore, the eigenvelue of
$(\hat{\phi}^*_i\hat{\phi}_j)$ is the number of particles \C{1}.

This physical interpretation allows the to introduce the correlations
in the ``bootstrap" manner considering the system under investigation
as the part of a ``big" system. In this case all parameters of the
$S$-matrix theory generating functional (\ref{1}) will be fixed.

C. {\it Perturbation theory}.

The eq.(\ref{4}) was derived in the perturbation theory framework.
The generalization of this formulae on the case of nonperturbate
contributions is the easy task considering fields on the real-time
contours $C_{\pm}$. One can use for this purpose the stationary
phase method if $Z(\phi_{\pm})$ are calculated, or the unitary formalism
\C{manj} if $R_0$ is calculated (last quantity is preferable if the
closed-path boundary conditions \C{1} shell be used).

But if the fields are defined on the time contour $C_{\b}(T_-,T_+)$,
whith both the real- and the imaginary-time parts, the
introduction of nonperturbative contributions is the hard problem
since it needs the definition of Green functions $G(t,t')$ with,
for instance, $t \in C_1$ and $t' \in C_2$.

D. {\it The equilibrium condition}.

Since $F(\b_+ ,\b_-)$ is the essentialy nonlinear function:
\be
F(\b_+ ,\b_-)=\sum^{\infty}_{n_+ ,n_- =0}
\frac{(\b_+ -\b )^{n_+}}{n_+ !}\frac{(\b_- -\b )^{n_-}}{n_- !}
F^{(n_+ ,n_-)}(\b ),
\l{}
\ee
the definition of integrals (\ref{8}) on the Gauss measure leads to
the asymptotic series. The coefficient of the expansion grows
\be
\sim \frac{\Gamma ((1+\sum^{\infty}_{k=1}k n_k )/2)}{n_3 ! n_4 ! \cdots},
\l{}
\ee
where the summation over all $n_k \geq 0$, $k=3,4,...$ is assumed.

The existence of this asymptotic series in the Borel sense depends
from the location of singularities over $\b$ of the $R_S (\b )$. We
had assume that this series exist by the following reason.

There is the general statement of statistical physics that the canonical
(Gibbs) and the microcanonical descriptions are equivalent (at least for the
equilibrium case). Following to this statement there are two possibilities.

(i). If the descussed asymptotic series can not be defined then
we must postulate that
\be
F^{(n_+ ,n_-)}(\b ) \equiv \frac{\partial^{n_+ +n_-}}{\partial^{n_+}\b_+
\partial^{n_-} \b_- }F(\b_+ ,\b_-))|_{\b_{\pm}=\b} \equiv 0,
n_+ +n_- \geq 3
\l{}
\ee
since in opposite case the canonical and the microcanonical descriptions
should not coinside. But it is the too strong constraint for interacting
fields.

(ii). If the asymptotic series can be ``regularised" in the Borel
sense than it is sufficient to assume the asymptotic condition:
\be
F^{(n_+ ,n_-)}(\b ) \sim 0, n_+ +n_- \geq 3
\l{14}
\ee
It is weaker constraint which also guaranties the Gaussian fluctuations
of $\b_{\pm}$. Inserting (\ref{15}) into (\ref{14}) one can easely
find that (\ref{14}) means smallness of $(n_+ +n_-) \geq 3$-point
energy correlation functions in comparision with 2-point energy
correlation function. Note that the condition (\ref{14}) leads to
the well known Boltzman's two-particle approximation. This formal
remark can be used as the support of above mentioned assumption that
the asymptotic series exist in the Borel sense.

Note that the constraints (\ref{14}) can be measured experimantaly.
It is important in the particles physics for investigation of the
created quark-gluon plasma state.

\vspace {0.2in}
{\Large \bf Acknowledgement}
I would like to thank Prof.~Mohanthappa for stimulating remark concernning
the closed-time path formalism.


\vspace {0.2in}

\begin {thebibliography}{99}

\bibitem{1}J.Manjavidze,{\it Preprint,\bf IP GAS-HE-5/95,
(hep-ph/9506424)}(1995)
\bibitem{mats}T.Matsubara,{\it Prog.Theor.Phys.\bf 14},351(1955)
\bibitem{2}J.Manjavidze,{\it Preprint,\bf IP GAS-HE-6/95 (hep-ph/9510251)}
(1995)
\bibitem{wig}E.P.Wigner,{\it Phys.Rev.,\bf 40}, 749 (1932)
\bibitem{car}P.Carrusers and F.Zachariasen, {\it Pfys.Rev., \bf D13},
950 (1986)
\bibitem{mil}R.Mills,{\it Propagators for Many-Particles Systems}
(Gordon and Breach, Science, 1970)
\bibitem{moh}P.M.Bakshi and K.T.Mohanthappa, {\it Journ.Math.Phys.,
\bf 4},1 (1961), {\it ibid, \bf 4}, 12 (1961)
\bibitem{sem}A.J.Niemi and G.Semenoff,{\it Ann.Phys.(N.Y.), \bf 152},
105 (1984)
\bibitem{*}H.Matsumoto, Y.Nakano and H.Umetzava, {\it J.Math.Phys.,\bf 25},
3076 (1984)
\bibitem{kms}R.Kubo,{\it J.Phys.Soc.Jap., \bf 12}, 570 (1957),
M.Martin and J.Schwinger, {\it Phys.Rev., \bf 115}, 342 (1959)
\bibitem{sch}J.Schwinger, {\it Particles, Sources and Fields \bf Vol.1}
(Addison-Wesley Pabl.Comp., 1970)
\bibitem{hu}E.Calsetta and B.L.Hu, {\it Phys.Rev., \bf D37}, 2878 (1988)
\bibitem{manj}J.Manjavidze, {\it Preprint, \bf IP GAS-HE-7/95
(quant-ph/9507003)}(1995)

\end{thebibliography}
\end{document}